\documentstyle[prl,aps,12pt,epsf]{revtex}
\def\inseps#1#2{\def\epsfsize##1##2{#2##1} \centerline{\epsfbox{#1}}}
\title{Collapse transitions of a periodic hydrophilic hydrophobic chain}
\author{E. Orlandini and T. Garel \\
Service de Physique Th\'eorique\\
CE-Saclay, 91191 Gif-sur-Yvette Cedex, France}

\date{\today}

\def\be{\begin{equation}}
\def\th0{\theta_0}

\def\bea{\begin{eqnarray}}
\def\ee{\end{equation}} 
\def\eea{\end{eqnarray}} 

\begin{document}
\maketitle

\vskip 125mm
\noindent\mbox{Accepted for publication in:} \hfill 
\mbox{Saclay, SPhT/98-022}\\ \noindent \mbox{``Eur. Phys. J. B''}\\ 
\vskip 1cm\noindent \mbox{PACS: 61.41.+e; 87.15.-v; 64.70.-p} \par
\noindent{Short title: Transitions of a periodic hydrophilic
hydrophobic chain}
\newpage
\begin{abstract}
We study a single self avoiding hydrophilic hydrophobic polymer
chain, through Monte Carlo lattice simulations. The affinity of
monomer $i$ for water is characterized by a
(scalar) charge $\lambda_{i}$, and the monomer-water interaction
is short-ranged. Assuming incompressibility yields
an effective short ranged interaction between monomer pairs $(i,j)$,
proportional to $(\lambda_i+\lambda_j)$. In this article, we take
$\lambda_i=+1$ (resp. ($\lambda_i=- 1$)) for hydrophilic
(resp. hydrophobic) monomers and consider a chain with (i) an equal
number of hydro-philic and -phobic monomers (ii) a periodic distribution of
the $\lambda_{i}$ along the chain, with periodicity $2p$. The
simulations are done for various chain lengths $N$, in $d=2$ (square
lattice) and $d=3$ (cubic lattice). There is a critical value $p_c(d,N)$ of
the periodicity, which distinguishes between different low temperature
structures. For $p >p_c$, the ground state corresponds to a 
macroscopic phase separation between a dense hydrophobic core and
hydrophilic loops. For $p <p_c$ (but not too small), one gets a microscopic
(finite scale) phase separation, and the ground state corresponds to
a chain or network of hydrophobic droplets, coated by hydrophilic
monomers.  We restrict our study to two extreme cases, $p \sim  O(N)$
and $p\sim O(1)$ to illustrate the physics of the various phase
transitions. A tentative variational approach is also presented. 
\end{abstract}

\newpage
\section{Introduction}
\label{sec: intro}
A very popular approach to the protein folding problem, is to
emphasize the heterogeneity of a protein due to the different side
chains \cite{Ga}. Of special importance in this 
context is the model of a (quenched) random
hydrophilic hydrophobic chain, since (i) proteins are usually designed to
work in water (ii) the first step of the folding transition may
correspond to the formation of a hydrophobic core \cite{Ag_Sh_Ud}. In
this model, each monomer is described by a single ``charge''
$\lambda_i$, and the polymer-water interactions are modelled through
the Hamiltonian:    
\begin{equation}
\label{diso1}
{\cal H}_{pw} = - \sum_{i=1}^{N} \sum_{\alpha=1}^{{\cal N}}  \lambda_i
\ a ( {\vec r_i}-{\vec R_{\alpha}} ) 
\end{equation}
where $a(\vec r)$ denotes a short-ranged (Van der Waals-like)
monomer-water molecule 
interaction, $N$ and $\cal N$ denote respectively the number of
monomers and of water molecules, and ${\vec r_i}$ and ${\vec
R_{\alpha}}$, their respective positions. Assuming that the system is
incompressible, one gets \cite{Ga,Obu}  
\be
\label{diso2}
{\cal H}_{pw} = + {1 \over 2} \sum_{i=1}^{N} \sum_{j=1}^{N}
(\lambda_i+\lambda_j) \ a( {\vec
r_i}-{\vec r_{j}}) - A \sum_{i=1}^{N} \lambda_i
\end{equation}
where $A= \sum_{\vec r} a(\vec r)$. The second term in (\ref {diso2})
is a constant for the quenched disordered chain (as well as for the
periodic chain studied in this article). It will therefore be
omitted henceforth.  
The phase diagram of this random
hydrophilic hydrophobic polymer has been studied by various methods
(mean-field, dynamics, replica variational calculations,....) for simple
disorder distributions of the $\lambda_i$ 's
\cite{Ga_Le_Or,Thi_Ash_Bhat,Mo_Ku_Da,Mo_Ku_Da2}. The results can be
roughly summarized as follows \par
(i) for strongly (on average) hydrophobic chains, one
expects a $\theta$ collapse transition to occur first, followed at
lower temperatures, by a scale dependent freezing transition.\par
(ii) for chains that are (on average), either weakly
hydrophobic or hydrophilic, one expects a freezing 
transition that is perhaps characterized by a ``random first
order''transition \cite{Ki_Thi_Wo}. 

An interesting application of this model to proteins has recently been
published \cite{Ha_Ta_Wi}, even though their hydrophobic content
appears not to be random \cite{Ir_Pe_Ro}.\par

As a first step towards the protein folding problem, we will study
numerically a single self avoiding chain, where the 
distribution of the $\lambda_i$ 's is periodic along the chain. To be
more specific, the chain of $N$ monomers is made out of 
periodically alternating blocks of $p$ hydrophilic monomers
($\lambda_i=+1$) and $p$ hydrophobic monomers ($\lambda_i=-1$).  Following equation (\ref{diso2}), the Hamiltonian of the chain
is defined as 
\be
\label{diso3}
{\cal H} = + {1 \over 2} \sum_{i=1}^{N} \sum_{j=1}^{N}
(\lambda_i+\lambda_j) \ \Delta( {\vec
r_i}-{\vec r_{j}})
\end{equation}
where $\Delta( {\vec
r_i}-{\vec r_{j}})$ denotes a lattice $\Delta$ function: $\Delta( {\vec
r})=1$, if  ${\vec r}$ connects nearest neighbours, and $0$ otherwise.
In this paper, we will consider square ($d=2$) and cubic ($d=3$) lattices.

The parameter $p$ is an important ingredient of the problem since
there is a critical value of $p$, that we note 
$p_c(d,N)$, above which
the equilibrium ground state consists of a single hydrophobic
core. The argument is as follows; for a spherical core ($d=3$), the number
of points on the surface is of order $N^{{2 \over 3}}$, whereas the
surrounding hydrophilic loops 
occupy a number of surface points of order $N \over p$. Therefore, one
expects a single macroscopic hydrophobic core for $p >O(N^{1 \over 3})$,
and many microscopic hydrophobic droplets for $p <O(N^{1 
\over 3})$. We therefore get $p_c(3,N) \sim O(N^{1 \over 3})$. For
$d=2$, the steric constraints are much stronger, and one 
hydrophilic loop anchored on the hydrophobic core screens part of its
``surface''. Since the distance between the two ends of
the hydrophilic loop typically scales with the radius of gyration of
the core, that is also with its ``surface'', this screening effect is
very strong. It implies that $p_c(2,N) \sim O(N)$. 
In the case $p < p_c(d,N)$, corresponding to many hydrophobic
droplets, one should further distinguish between a linear or branched  
topology of the hydrophobic droplets. For very small $p$,
the very formation of these droplets is impeded by the hydrophilic
parts of the chain (see below)

The numerical simulations for the periodic chain of equation
(\ref{diso3}) are done on the square or cubic lattice, using the
Multiple Markov chain method \cite{Or}. The plan 
of the paper is the following. We first recall, in section
\ref{sec:numer} the principles of the Multiple Markov chain method in
Monte Carlo simulations of polymer chains. As a preliminary test, we
apply this method to the two dimensional $\theta$ collapse of a purely
hydrophobic chain. We then consider the periodic hydrophilic
hydrophobic chain in two extreme cases, namely $p \sim O(N)$ and $p \sim
O(1)$, to capture  the physics of the various phase
transitions. Putting numbers on the above estimates of $p_c$  shows
that it will be numerically convenient to study the single core
phase in $d=3$ and the multiple cores' phase in
$d=2$. In section \ref{sec:single}, the case $p \sim O(N)$ is studied
in detail for $d=3$; we also briefly consider the corresponding transition
in $d=2$.  In section \ref{sec:many}, we study the case $p\sim
O(1)$ which, as stated above, has a multiple droplet low temperature
structure: for $d=2$, we present evidence for the existence of an
intermediate branched phase, if $p$ is not too small. The $d=3$ case
requires very long chains and is only briefly considered. Finally, we also 
present a tentative variational method for this case.

\section{The Multiple Markov chain method}
\label{sec:numer}
\subsection{Summary of the method}
In this section we give a quick description of the numerical
techniques which we use to calculate thermodynamic (and/or geometric)
properties of the chains as a 
function of the temperature. These techniques have been discussed in
detail in reference \cite{TJOW}, and can be summarized as follows. 

The implementation of the Metropolis Monte Carlo method relies on the
multiple Markov chain sampling. First,
one generates with the simple Metropolis heat bath, a Markov chain at
temperature $T$: this procedure makes use of an hybrid algorithm, which
has pivot \cite{MS87}  as well as crankshaft moves \cite{VS}.
Pivot moves are of a global type, and operate well in the swollen phase;
crankshaft moves, which are of a local type, are essential
in speeding up convergence close to the collapse phase transitions
\cite{TJOW}. In these calculations, each
Monte Carlo step consists of $O(1)$ pivot move and $O(N)$ crankshaft
moves. 

One may then run in parallel a number $m$ (typically $15-20$) of these
Markov chains at different temperatures $T_1 > T_2 > \ldots > T_m$. We
allow the chains to interact (by possibly swapping conformations) as
follows. Among the $m$ chains, we select, with uniform probability,
two chains $({\alpha,\beta})$  at respective temperature $T_{\alpha}$ and
$T_{\beta}$. A trial move is an attempt to swap the two current
conformations of these chains.  If $\pi_K (T)$ is the probability of
getting state $K$ at temperature $T$ (that is $\pi_K (T) \sim
e^{-{H(K) \over T}}$, in 
obvious notations), and $S_{\alpha}$ and $S_{\beta}$ are the current
states in the $\alpha$-th and $\beta$-th chain, then we accept the
trial move (i.e. we swap $S_{\alpha}$ and $S_{\beta}$) with
probability   
\begin{equation}
r(S_{\alpha} , S_{\beta}) = \min \left( 1, \frac{\pi_{S_{\beta}} (T_{\alpha})
\pi_{S_{\alpha}} (T_{\beta})}{\pi_{S_{\alpha}} (T_{\alpha})
\pi_{S_{\beta}} (T_{\beta})}  
\right) .
\label{MCC}
\end{equation}
Note that the whole process is itself a (composite) Markov chain.  Since the
underlying Markov chains are ergodic, 
so is the composite Markov chain. Furthermore, the composite
chain obeys detailed
balance, since the swap moves as well as the moves in the underlying
chains obey detailed balance \cite{TJOW}. The swapping procedure
dramatically decreases the correlations within each Markov chain, and
produces little CPU waste since, in any case, one is interested in
obtaining data at many temperatures\cite{Or}. 

\subsection{The two dimensional $\theta$ transition}
\label{2theta}
We first consider the application of the Multiple Markov chain method
to the two dimensional $\theta$ collapse
transition of a purely hydrophobic chain (see ref \cite{TJOW} for the
case $d=3$). We are mostly interested here in locating the
thermodynamic $\theta$ transition.
\par
We show in Figure 1 the plot of the average scaled radius of gyration
$({<R^{2}> \over N^{2\nu_{\theta}}})$ versus temperature, for
different values of $N$. Finite size scaling theory
\cite{Mi_Pa_Bi,Wi_Kr_Go,Bau,Mei_Lim,Gra_Heg,Gra_Heg2} predicts that
\begin{equation}
\label{fss}
{<R^{2}> \over N^{2\nu_{\theta}}}=f(tN^{\Phi})
\end{equation}
where $t=\vert {T-T_{\theta} \over T_{\theta}} \vert$ is the reduced
temperature for $N$ large ($T_{\theta}=T_{\theta}(N=\infty)$), $\Phi$
a crossover exponent, and $f(x)$ a scaling 
function with a finite value for $x \to 0$. Using the exact result
$2\nu_{\theta} ={8 \over 7}$ \cite{Dup_Sal} yields an estimate of the
$\theta$ temperature: $T_{\theta} \simeq 1.5$, in agreement with
recent estimates \cite{Gra_Heg}. To study the critical 
behaviour of the specific heat close to the 
$\theta$ point is a notoriously difficult problem in $d=2$. We
have also verified this 
point, and we have only extracted from the $N$ dependence of the peak of the
specific heat (Figure 2) a large $N$  critical temperature
$T'_{\theta} \sim 1.5$ in broad agreement with the value obtained
from the behaviour of the radius of gyration.

\section{The case p =O(N)}
\label{sec:single}
\subsection{Numerical simulations for d=3} 
\label{sec:num3}
This case corresponds to a single hydrophobic core ground
state since $p_c(3,N) \sim O(N^{1 \over 3})$. In the case of the
$\theta$ collapse transition, the $N$ monomers of the chain
play an equivalent role (neglecting end effects for a long enough
chain). In the present problem, the repulsive interaction between
hydrophilic monomers leads us to consider two possibilities (i) 
the collapse transition first occurs in a single hydrophobic block of
length $p$ (ii) the collapse is due to a cooperative effect of the ${N
\over 2p}$ 
hydrophobic blocks, in a way similar to the $\theta$
transition. Scenario (ii) is consistent with a unique phase transition of a
discontinuous character occurring at temperature $T_{\theta}({N \over
2})$. Scenario (i) is a priori consistent with a collapse
transition at $T_{\theta}(p)$, and raises the question of a (surface
induced) sticking transition of the individual hydrophobic blocks.

In either case, the collapse transition is expected to be
discontinuous, with a jump of the radius of gyration. 
We have fixed $p={N \over 8}$ in the simulations, and let $N$ vary from
80 to 640.  A typical low temperature configuration for $N=640$ is
shown in Figure 3: as expected, it displays a macroscopic phase
separation between the hydrophobic and hydrophilic parts of the
chain. To get an estimate of the critical temperature, we have first
considered the radius of gyration $<R^2>$ of the complete chain.
As shown in Figure 4, 
the exponent of $<R^2>$ is, at all 
temperatures, given  by the self avoiding walk
(SAW) value ($\nu_{SAW} \simeq 0.588$ \cite{Leg_Zin,Gui_Zin}). 

Our data for the large values of $N$ are in agreement with scenario
(i), that is a single hydrophobic block collapse, since the critical
temperature $T_c$ is very close to $T_{\theta}(p)$ (see Table I).
A better estimate of $T_c$ comes from specific heat measurements,
since the specific heat has a quite spiky character (Figure 5), unlike
its $\theta$ point equivalent \cite{TJOW}. We again get an estimate
very close to $T_{\theta}(p)$. 
We have also tried a finite size scaling analysis to find $T_c$, for
large $N$ (and therefore large $p$). Following a well
established path (see e.g.\cite{Muthu} and references therein), we
have considered the $N$ dependence of the height and position of the
peak in the specific heat.  At the transition, and in the scaling
limit, one expects the height peak $C^{*}$ to scale like 
\begin{equation} 
\label{spec}
C^{*} \sim N^{2 \Phi -1} 
\end{equation}
where the crossover exponent $\Phi$ and the specific heat exponent
$\alpha$ are related \cite{La} through the relation $\Phi
(2-\alpha)=1$. At the three dimensional $\theta$ transition, one has
$\Phi=0.5$ and $\alpha=0$. At a (thermal) first order transition, one
has $\alpha= 1$, yielding $C^{*} \sim N$. Finite size scaling also
implies a critical temperature shift $\Delta T_c=T_c(\infty)-T_c(N)
\sim {1 \over  N^{\Phi}}$, yielding $\Delta T_c \sim {1 \over N}$
for a first order transition. Figures  6(a) and 6(b) show our results,
and confirm that our simulations are not done in the scaling regime.
It is well known indeed that,
in order to get a precise estimate of the thermodynamic $\theta$
temperature, one has to study very long chains (typically $N >
1500$). For our problem, since $p={N \over 8}$, we should study the
case $N=12000$, which is presently out of reach. We have done an
independent simulation with $N=800$, and $p={N \over 4}=200$. The peak
in the specific heat occurs for $T_c \simeq 2.5$, very close with
$T_{\theta}(p=200)$ (see Table I). 

We therefore believe that, in the
thermodynamic limit ($N,p \to \infty$), one has a discontinuous single
block collapse transition; the critical temperature $T_c({\infty})$ is
the same as the collapse temperature of the fully hydrophobic chain
($T_{\theta}=T_{\theta}(\infty) \simeq 3.7$ \cite{Gra_Heg2}). The
transition is well characterized by the phase separation order
parameter \cite{Mo_Ku_Da,Mo_Ku_Da2}:   
\begin{equation}
\label{phasesep}
<\delta R^2>=<R^2_{phil}>-<R^2_{phob}>
\end{equation}
where $<R^2_{phil}>$ (resp. $<R^2_{phob}>$ is the squared radius of
gyration of the hydrophilic (resp. hydrophobic) monomers.
In Figure 7, we have
plotted the order parameter  $({<\delta R^2> \over N^{2 \nu_{SAW}}})$ as
a function of temperature, for various values of $N$. Its behaviour is
consistent with the previous results.

An intriguing feature of the single block collapse mechanism is that
it seems to imply the existence of an intermediate phase, between the
low temperature phase depicted in Figure 3, and the swollen coil
phase.  This intermediate phase is a necklace (or
network) of single hydrophobic blocks, and its free energy differs
from the low temperature free energy by a surface free
energy.  We have not found this intermediate phase in our multiple 
chain Monte Carlo Method, and this may have several causes. It is for
instance possible that the temperatures of the
various chains $T_1 > T_2 > \ldots > T_m$ of our simulations had too
large a spacing to find this phase. Further work, in particular a
precise determination of surface properties, is needed on this point.


\subsection{Numerical simulations for d=2}
\label{sec:num2}
In this case, one expects $p_c(2,N) \sim O(N)$, yielding two different
situations. If $p>p_c$, one will get a single hydrophobic core below
the  collapse transition (Figure 8).  One may then study the (full)
radius of gyration at high and low temperatures (Figure 9), obtaining
in both cases self avoiding behaviour (i.e. $\nu={3 \over 4}$
\cite{Nie}). If $p<p_c$, several hydrophobic 
cores appear (Figure 10). On the square lattice, we have found
numerically that $0.06 <p_c(2,N)/N<0.08$. One may get a feeling for
this result by considering a square shaped hydrophobic single core
that is fully surrounded by hydrophilic loops: the above estimates
corresponds to a total number of hydrophilic loops approximately equal
to $6-8$. This argument is only suggestive since the single core
becomes elongated as $p \to p_c^{+}$. This is clearly 
due to the screening 
effect of the hydrophilic loops: the hydrophobic core tries to maximize
its perimeter at fixed surface (Figure 11). As for the phase
transition for $p>p_c$, one may say that specific heat data display a
rather smooth behaviour; the transition is not markedly
discontinuous. As already mentionned, extracting a more detailed
information from these data is rather tricky in two dimensions.
\section{The many hydrophobic droplet chain p=O(1)}
\label{sec:many}
\subsection{Numerical simulations for d=2}
\label{sec:num5}
The large value of $p_c(2,N)$ shows that the multiple droplets' phase
should be a priori easier to study in two dimensions. We show in Figure
12 a typical low temperature configuration, which displays
branched polymer features \cite{Lu_Is}. From section 
\ref{sec:num2}, we know that the maximum number of hydrophobic
monomers in a droplet is $n_{MAX} \sim 12-16 \ p$. To further investigate the 
branched character of the phase, we have studied the case
$p=8$, with $N$ ranging from $80$ to $1200$ (other values of $p$ are
briefly considered below).  For $p=8$, the high and low  
temperature exponents $\nu$ of the radius of gyration are shown on Figure
13.  Above the
transition, we get SAW behaviour; below the 
transition, we get $\nu \simeq 0.64$, which is indeed close to the
branched polymer (BP) value \cite{De_De}. 

 It turns out that low temperatures
are difficult to study because of non-equilibrium effects, so that we
are not able to follow in detail the thermal evolution of the branched
phase. This is partially due to the fact that the Monte Carlo method
of section \ref{sec:numer} has not been optimized to deal with
branched phases. Another reason may be the possible existence of a
dynamical phase transition towards some kind of glassy branched state
(see section \ref{Variat}). 
We have therefore restricted our study to the
phase transition between the high temperature (SAW) and low temperature (BP)
phases. Using the same finite size argument as in section
\ref{2theta}, we plot in Figure 14 the scaled radius of gyration
$({<R^{2}> \over N^{2\nu^{*}}})$ vs temperature. Various values of
the unknown exponent $\nu^{*}$ have been considered (Figure 15). Our
results show the existence of a phase transition at $T_c \simeq 0.8
\pm 0.1$, and
evidence for a new critical behaviour $(\nu^{*} \simeq 
0.70 \pm 0.03)$ at $T_c$. The phase 
transition seems also to be quite smooth, if one considers specific
heat data. Another ``experimental'' observation concerns the size
distribution of the hydrophobic droplets: below the transition, we
find that most of the droplets do not reach the maximum size allowed
$n_{MAX}$. This can be interpreted as an entropic effect, very much along the
lines of reference \cite{Ki_Thi_Wo}. 

We conclude this section by a few remarks on the role of $p$. We have
also considered the case $p=4$, and $p=10,12$, with the same range of
values of $N$. We do not find a clear evidence for a phase transition
for $p=4$, whereas we find evidence for two transitions for
$p=10,12$. In the latter case, the branched phase gives way at low
temperature to a reentrant self avoiding chain of finite hydrophobic
droplets. This shows that the balance between linear and branched
topologies is very dependent on the value of $p$. If $p$ is too small,
the formation of the droplets is impeded by the repulsive hydrophilic
monomers. For large $p$, a local collapse is possible, favoring the
linear topology at low temperature. For intermediate values of $p$, a
non local hydrophobic collapse is apparently favored, yielding a
branched topology. These issues will be further
tackled in section \ref{Variat}.
\subsection{Numerical simulations for d=3}
\label{sec:num4}
A typical low temperature configuration is shown in Figure 16, for
$p=4, N=720$. The properties of the multiple hydrophobic cores phase
are difficult to study, since one needs very large values of
$N$. Furthermore, for $d=3$, a SAW at the $\theta$ point and a
branched polymer (BP) have the same exponent $\nu_{\theta}=\nu_{BP}=0.5$  
\cite{Pa_Sou}, which makes a detailed scaling
analysis difficult. 


\subsection{Variational method for the many droplet phase}
\label{Variat}
\par
Following traditional polymer physics methods \cite{Ga}, we will study the low
temperature branched phase in a variational way. The basic
steps can be summarized as follows:  \par 
(i) one derives an effective quantum Hamiltonian. 
\par
(ii) ones uses a ground state approximation, together with a saddle
point approximation. 
 \par
(iii) finally, one performs a variational calculation, and minimizes the
free energy with respect to the relevant parameters.\par

Steps (i) and (ii) are familiar in the context of the usual $\theta$
collapse transition. Since one is then interested in a macroscopic
collapse, a continuous description of the chain is valid; furthermore,
the ground state approximation holds for long enough chains, since
there is a bound state representative of the 
collapsed globule. Finally, step (iii) is usually implemented with a constant
or Gaussian density around the center of mass of the collapsed
globule, which can be taken as fixed in all calculations. 
\par
On the contrary, what we have in the low temperature branched
phase is a inhomogeneous liquid of microscopic hydrophobic
droplets. If one follows the above procedure, one must take the
extensive entropy of these droplets (i.e. the degeneracy of the saddle
point) into account. Since we believe that our approach may be of
interest in other contexts \cite {Ki_Thi_Wo}, we will assume that a
continuum description of the chain is valid, and derive the simplest
form of the associated Hamiltonian. We will also assume that ground
state dominance holds.  

The partition function of the hydrophilic hydrophobic chain reads
\be
\label{partit1}
Z = \int \prod_{i}  d{\vec r_i} \   e^{-{d \over 2a^{2}}\sum_{i}(\vec r_{i+1}-\vec
r_{i})^{2} - \beta {\cal H}}  
\ee
 where $a$ is a typical monomer length and $i=1,2,...N$. The lattice
Hamiltonian (${\cal H}$) has been derived in equation (\ref{diso3}), 
and its off-lattice version reads
\bea
\label{hamhyd}
\beta {\cal H} =&& {1 \over 2} \sum_{i \ne j} \left [
v_0 + \beta ( \lambda_i+\lambda_j)\right]
 \ \delta(\vec r_i - \vec r_j) \nonumber\\
&&+ { 1 \over
6 } \sum_{i \ne j \ne k} w_0
\ \delta(\vec r_i- \vec r_j)\ \delta(\vec r_j - \vec r_k) \nonumber\\
&&+ { 1 \over 24 } \sum_{i \ne j \ne k \ne l} y_0
\ \delta(\vec r_i- \vec r_j)\ \delta(\vec r_j - \vec r_k)
\ \delta(\vec r_k - \vec r_l)
\eea
Note that we have also included three ($w_0$) and four ($y_0$) body
terms for reasons that will soon become clear. We also assume ($\vec
r_1=\vec r_N =\vec 0$). \par 
Defining the local density $\rho(\vec r)$ as
\bea
\label{density}
\rho(\vec r)&=& \sum_{i} \ \delta (\vec r- \vec r_i) 
\eea
we have
\be
\label{partit2}
Z =   \int {\cal D} \phi ( \vec r ){\cal D} \rho ( \vec r )
 \ \zeta (\rho,\phi) \ \exp \left ( i \int d^d r \phi ( \vec r ) \rho ( \vec r ) 
-  \int d^d r \left[{v_0 \over 2} \rho^2 (\vec r)+ {w_0 \over 6}
\rho^3 (\vec r) +{y_0 \over 24} \rho^4 (\vec r)\right] \right ) 
\ee
with
\be
\label{partit3}
\zeta (\rho,\phi)= \int \prod_{i}  d{\vec r_i} \   e^{-{d \over 2a^{2}}\sum_{i}(\vec r_{i+1}-\vec
r_{i})^{2} - i\sum_i \phi(\vec r_i) -\beta \sum_i \lambda_i \rho(\vec r_i)}  
\ee
For the periodic hydrophilic hydrophobic chain, we introduce transfer
matrices $T_{\pm}$ for $\lambda_i=\pm1$ and get
\be
\label{partit4}
\zeta(\rho, \phi)=<\vec 0| \left[ T^{p}_{+}T^{p}_{-}\right]^{N \over
2p} |\vec 0>  
\ee
with 
\be
\label{transfer1}
<\vec r| T_{+} | \vec r'>= e^{-{d \over 2a^{2}}(\vec r-\vec r')^{2} - i \phi(\vec r) -\beta \rho(\vec r)}
\ee
and
\be
\label{transfer2}
<\vec r| T_{-} | \vec r'>= e^{-{d \over 2a^{2}}(\vec r-\vec r')^{2} - i\phi(\vec r) +\beta \rho(\vec r)}
\ee
In equation (\ref{partit4}), the differences between $p \sim O(N)$ and
$p \sim O(1)$ are clearly displayed. From now on, we will set $p=1$,
without questioning any further the existence of the continuum limit
in this case. Using the identity 
\be
\label{ident1}
<\vec r| T_{+}T_{-} | \vec r'>=\int d^d r_{0} <\vec r| T_{+} | \vec
r_0><\vec r_0| T_{+} | \vec r'> 
\ee
together with equations (\ref{transfer1})(\ref{transfer2}) leads, to
lowest non trivial order in $\phi(\vec r)$ and $\rho(\vec r)$, to
\be
\label{transfer3}
<\vec r| T_{+}T_{-} | \vec r'>=e^{-{d \over 4a^{2}}(\vec r-\vec
r')^{2} - 2i \phi(\vec r) + {a^{2} \over 4d} \beta^{2} {({\vec
\nabla}\rho)}^{2}(\vec r)} 
\ee
which implies
\be
\label{zeta1}
\zeta(\rho, \phi)= {\rm Tr} \ \exp {-({N \over 2} {\cal H}_0)}
\ee
with 
\be
\label{quant1}
{\cal H}_0=  -{a^2 \over 4d} {\vec \nabla}^2 + 2i \phi(\vec r)- {a^{2}
\over 4d} \beta^{2} ({\vec  \nabla}\rho)^{2}(\vec r)  
\ee
 It is quite clear that our derivation is not rigorous. We
nevertheless feel that the $\rho (\vec r)$ dependent term of the Hamiltonian
${\cal H}_0$ is physically sound since it favors inhomogeneous high
density regions (droplets). A better approximation would presumably
involve higher derivatives of $\rho(\vec r)$, which clearly define
typical droplet sizes. Assuming ground state
dominance in (\ref{quant1}), and performing a saddle point approximation
on $\phi(\vec r)$ in (\ref{partit2}) yields
\begin{equation}
\label{psi2}
\rho(\vec r)=N \Psi^{2}(\vec r)
\ee
where $\Psi(\vec r)$ is a normalized wave function. We then obtain the
``free energy''per monomer as  
\be
\label{f1}
\beta {F \over N}=\min_{\{\Psi(\vec r)\}} \int d^{d} r \ G({\Psi(\vec r)})
\ee
where
\bea
\label{variat1}
 G(\Psi(\vec r))=&& \int d^{d} r
\left({v_0 N \over 2} \Psi^{4}+{w_{0} 
N^2 \over 6} \Psi^6 + {y_0 N^3 \over 24} \Psi^8 \right) \nonumber\\
&&+ {1 \over 2}  \int { d}^d r  \Psi(\vec r) \left(-{a^2 \over 4 d }
{\vec \nabla}^2 - {a^{2} \over 4d} \beta^{2} ({\vec  \nabla} (N
\Psi^{2}))^{2}(\vec r)\right) \Psi (\vec r)   
\eea
 At this point, it is useful to remark that one needs in this
approximation to introduce four body interactions, as in the
disordered case, and for the same reasons \cite{Ga_Le_Or}. The fact
that attractive multi-body interactions in homopolymers may induce a
SAW-BP phase transition has been previously noted for a specific model
in reference \cite{Or_Se_St_Te}.

At low temperature, the last term of equation (\ref{variat1}) plays a
dominant role. This term, as mentionned above, tends to create an
interface between hydrophilic and hydrophobic regions. If one uses
a variational wave function $\Psi_0(\vec r)$ given (for $d=2$) by
\be
\label{variat2}
\Psi_0(\vec r)=a_0 \ {\rm cos}({2 \pi x \over l}) \ {\rm cos}({2 \pi y \over l})
\ee
the normalization condition implies
\be
\label{norm}
a_0^{2}R^{2} \sim O(1)
\ee
where $R$ is the linear dimension of the system. Plugging this
estimate in equation (\ref{variat1}) shows that a low free energy is
obtained for $l \sim a$ and a finite average density $\rho_{av}={N
\over R^2} \sim O(1)$. In other words, the low temperature phase
obtained from our variational approach is a dense phase (since $\nu={1
\over d}$, with $d=2$), made of
microscopically phase separated regions. In section
(\ref{sec:num5}), we obtained from our simulations the value $\nu
\simeq 0.64$ below the the (SAW)-(BP)
phase transition. This result is not compatible with the value
$\nu=0.5$ that we get through the variational method. Some
possible explanations for this ``discrepancy'' 
are as follows

(i) The simulations were done for $p=8$ on a lattice, and the
variational method was applied to the case $p=1$, within a continuum
limit approach. 

(ii) A major difference between the present multi droplet collapse and the
$\theta$ single core collapse is that one has to take into account the
degeneracy of the saddle point equation for $\phi(\vec r)$ in evaluating the
``true'' free  energy. In other words, there is a droplet entropy that must be
considered. Sticking with the variational function of
equation (\ref{variat2}), it is easily seen that this entropy 
favors large values of ${l \over a}$. A precise calculation is
difficult, and we will not comment upon this point anymore.
 
(iii) We believe however that the main reason for the difference
between the exponents $\nu$ stems from the use of a ground state
approximation in estimating the right handside of equation
(\ref{zeta1}). As far as we know, this approximation, which relies on
the existence of a bound state in the Hamiltonian ${\cal H}_0$, works
well for dense (finite density) phases. It does not a priori describe
a branched polymer, which has a vanishing density. Physically, a dense
phase is not favorable because of the repulsion between the
hydrophilic monomers.

Altogether, our results seems to indicate that the finite $p$
chain may undergo zero ($p=4$), one ($p=8$) or two ($p=10,12$)
phase transitions. Ground state dominance is never a valid
approximation, since one deals with either linear ($\nu =0.75$) or
branched ($\nu \simeq 0.64$) structures. Reentrant
behaviour, similar to the one described for $p=10,12$, has been found
recently in related models 
\cite{Tro_Va_Ma,Le_Bo_Ha}. Another issue of
interest concerns a possible low temperature dynamical (glass)
transition, similar to the one described in reference \cite{Ki_Thi_Wo}.

\newpage
\section{Conclusion} 
\label{sec:conc}

We have studied a periodic hydrophilic
hydrophobic chain. An important ingredient of the physics of this
problem is the value of the period $2p$. The low temperature phase
consists of a single ($p>p_c(d,N)$) or multiple ($p<p_c(d,N)$) hydrophobic
core(s), where $p_c(3,N) \sim O(N^{1 \over 3})$ and $p_c(2,N) \sim
O(N)$.  Using Monte Carlo calculations, we have studied the case $p
\sim O(N)$ and indeed found a 
macroscopic phase separation between the two types of monomers in
$d=3$, and two possible regimes in $d=2$. The second case, ($p \sim
O(1)$) yields for both dimensions a low temperature phase,
consisting of a chain or network of microscopic hydrophobic droplets linked by
hydrophilic filaments. We have studied this phase numerically
in $d=2$. A connection with both the branched polymer chain
\cite{Lu_Is} and the random hydrophilic hydrophobic chain
\cite{Ga_Le_Or,Thi_Ash_Bhat,Mo_Ku_Da,Mo_Ku_Da2,Tro_Va_Ma} is physically
appealing and shows up in the tentative variational treatment of this
phase, as given in section \ref{Variat}. As for proteins, the
existence of periodic hydrophobicity patterns in secondary structures
\cite{Ir_Pe_Ro} 
suggests that our model may have some relevance in explaining the
typical size of single domain proteins ($N \sim 120-150$ residues).
Further work in these
directions is in progress.
\vskip 4mm
It is a pleasure to thank Henri Orland for fruitful discussions and
suggestions, and Bernard Derrida for interesting comments.

\newpage
\begin{center}
{\bf Figure captions}
\end{center}\par
{\bf Figure 1:} Scaled radius of gyration $({<R^{2}> \over
N^{2\nu_{\theta}}})$ vs temperature for
the purely hydrophobic chain in $d=2$, for $N=80 \ (\triangle), 160
\ (\Box), 240 \ (\Diamond), 480 \ (\times), 640 \ (+)$. A crossing occurs for
$T_{\theta} \simeq 1.5$.\par  
{\bf Figure 2:} Specific heat vs temperature, for the purely
hydrophobic chain in $d=2$, for the same values of $N$. The
extrapolated critical temperature is $T'_{\theta} \simeq 1.5$.\par
{\bf Figure 3:} Typical phase separated configuration ($d=3$, $p={N
\over 8}$, $N=640$). Circles denote hydrophobic monomers.\par 
{\bf Figure 4:}   Log-Log plot of $<R^2>$ vs $N$ at various
temperatures for ($d=3$, $p={N \over 8}$), and
$N=80, 160, 240, 360, 480, 640, 800$. The temperature is $T=\infty \
(\Box), 3.33 \ (\triangle), 2.0 \ (*)$,\par ,$1.0 \ (\Diamond), 0.5 \
(+)$. The upper and lower straight lines have slopes compatible with
self avoiding behaviour $2\nu_{SAW} \simeq 1.176$.\par 
{\bf Figure 5:} Specific heat vs temperature for ($d=3$, $p={N \over
8}$), and $N=80 \ ({\rm O}), 160 
\ (\Box)$,\par $240 \ (\Diamond), 360 \ (\times), 480 \ (+), 640 \
(\triangle)$. Note the increase of the peak as 
well as its shape, when $N$ increases. \par
{\bf Figure 6:} (a) Plot of the specific heat peak $C^{*}$ vs
$N$ for ($d=3$, $p={N \over 8}$), and $N=80, 160, 240, 360, 480, 640$
(b) Plot of the critical temperature vs ${1 \over N}$ for the same
parameters. Error bars correspond to one standard deviation.\par 
{\bf Figure 7:}  Scaled phase separation parameter (see equation
(\ref{phasesep})) vs temperature, for ($d=3$, $p={N \over 8}$), and
$N=80, 160, 240, 360, 480, 640$.\par  
{\bf Figure 8:} Low temperature phase separated configuration ($d=2$,
$p={N \over 8}$, $N=240$). Black triangles denote hydrophilic
monomers. Note the isotropic shape of the hydrophobic
core ($p>>p_c$).\par
{\bf Figure 9:}   Log-Log plot of $<R^2>$ vs $N$ at various
temperatures for ($d=2$, $p={N \over 8}$), and $N=80, 160, 240, 360,
480, 640, 800$. The temperature is $T=\infty \ (\Box), 1.4 \
(\triangle), 1.0 \ (*), 0.5 \ (\Diamond)$. The upper and
lower straight lines have slopes compatible with self avoiding
behaviour $2\nu_{SAW} =1.5$.\par 
{\bf Figure 10:} Low temperature multiple cores' configuration
displaying the screening effect of the hydrophilic loops ($d=2$, $p={N
\over 24}$, $N=240$). \par 
{\bf Figure 11:} Low temperature phase separated configuration ($d=2$,
$p={N \over 12} \simeq p_c+$, $N=240$). Note the elongated shape of the
hydrophobic core.\par
{\bf Figure 12:} Typical multiple cores' configuration ($d=2$,
$p=8$, $N=1200$). Black triangles denote hydrophilic
monomers.\par
{\bf Figure 13:} Log-Log plot of $<R^2>$ vs $N$ at various
temperatures for ($d=2$, $p={8}$), and $N=80, 160, 240, 360, 480, 640,
800, 1200$. The temperature is $T=\infty \ (\Box), 1.0 \
(\triangle), 0.7 \ (*)$,\par ,$0.5 \ (\Diamond), 0.33 \ (+)$.The upper
straight line has a slope $2\nu_{SAW} =1.5$, 
whereas the middle line has slope 
$2\nu_{BP}$, where $\nu_{BP} \simeq 0.64$ is the branched polymer (BP)
value. The bottom line corresponds to the collapsed value $2\nu=1$. \par
{\bf Figure 14:} Scaled radius of gyration $({<R^{2}> \over
N^{2\nu^{*}}})$ vs temperature for ($d=2$, $p=8$) and  $N=80
\ (\Box), 160 \ (\Diamond), 240 \ (\times), 480 \ (+), 640 \
(\triangle)$. A crossing occurs for $\nu^{*} \simeq 0.70$, yielding
$T_c \simeq 0.8$.\par
{\bf Figure 15:}  With the same data, no crossing occurs for (a)
$\nu^*=\nu_{SAW}=0.75$ (b) $\nu^*=\nu_{\theta}={4 \over 7}$ (c) $\nu^* =
\nu_{BP} \simeq 0.64$.\par
{\bf Figure 16:} Typical multiple cores' configuration ($d=3$, $p=4$,
$N=720$). Circles denote hydrophobic monomers.\par
\newpage
\begin{center}
{\bf Table caption}
\end{center}\par
 {\bf Table I:} Comparison between the critical temperature of the
hydrophilic hydrophobic chain $T_c=T_c(p,N)$, and the $\theta$
transition temperatures \cite{TJOW} of a fully hydrophobic chain of (i) $p$
monomers (ii) ${N \over 2}$ monomers. The first three lines correspond
to $p={N \over 8}$ ($N=80,640,800$). The last line corresponds to
$p={N \over 4}$ ($N=800$). The value $T_{\theta}(p=10)$ has been obtained using exact enumeration techniques.

\newpage
\begin{figure}
\vskip 1.truecm
\inseps{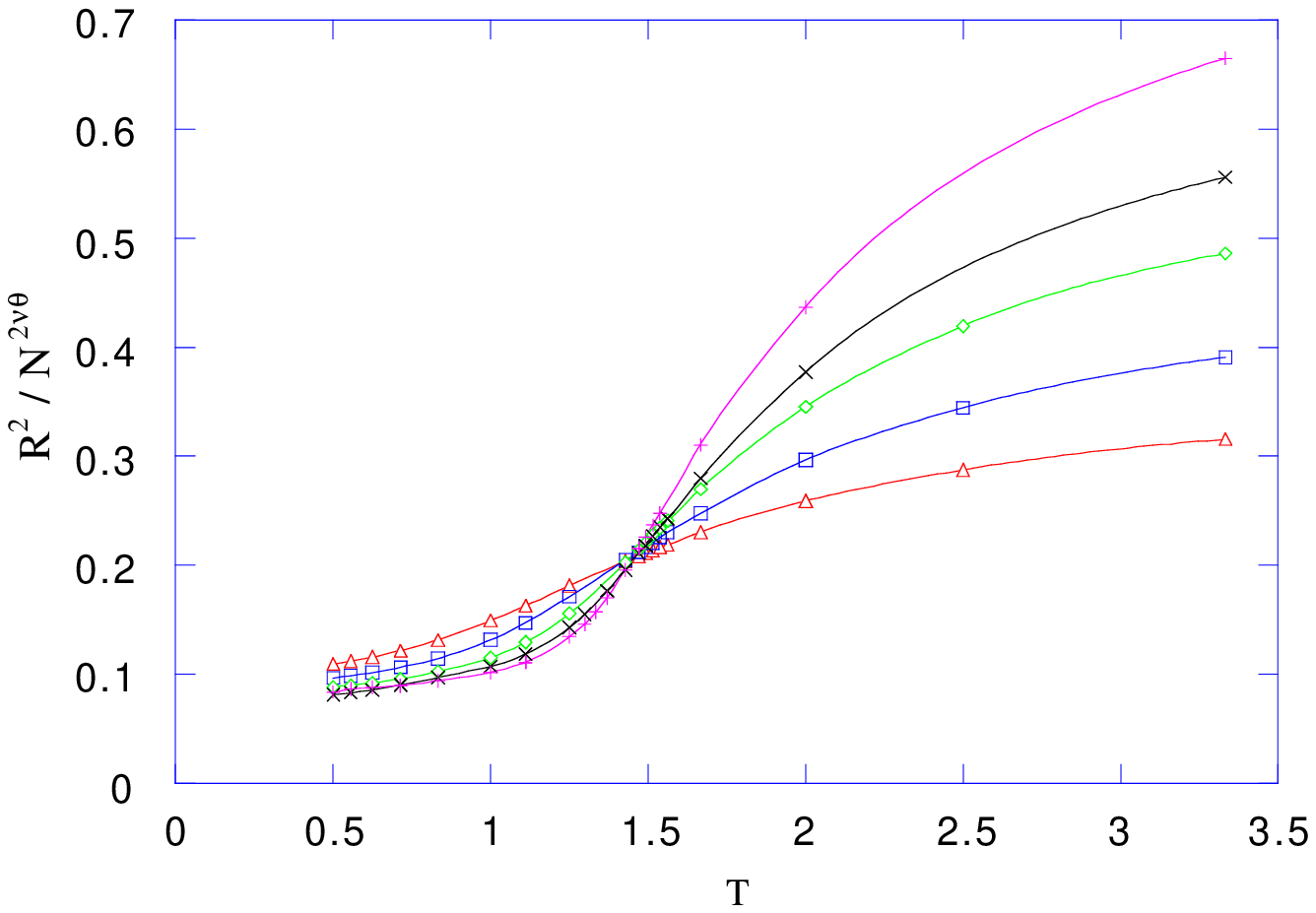}{1.2}
\vskip -3. truecm
\label{figure1}
\end{figure}

\begin{figure}
\vskip 1.truecm
\inseps{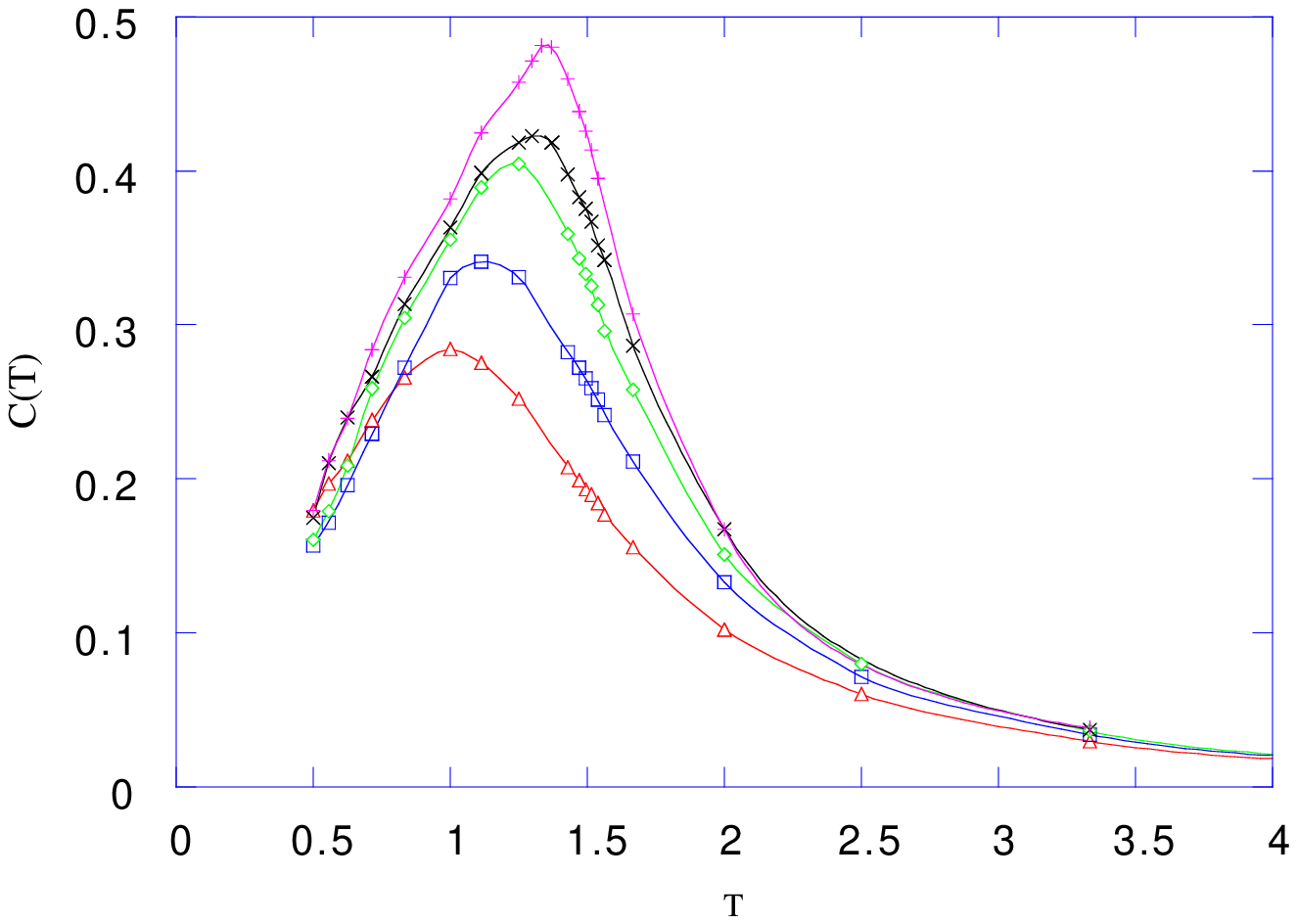}{1.2}
\vskip -1. truecm
\label{figure2}
\end{figure}

\begin{figure}
\vskip 1.truecm
\inseps{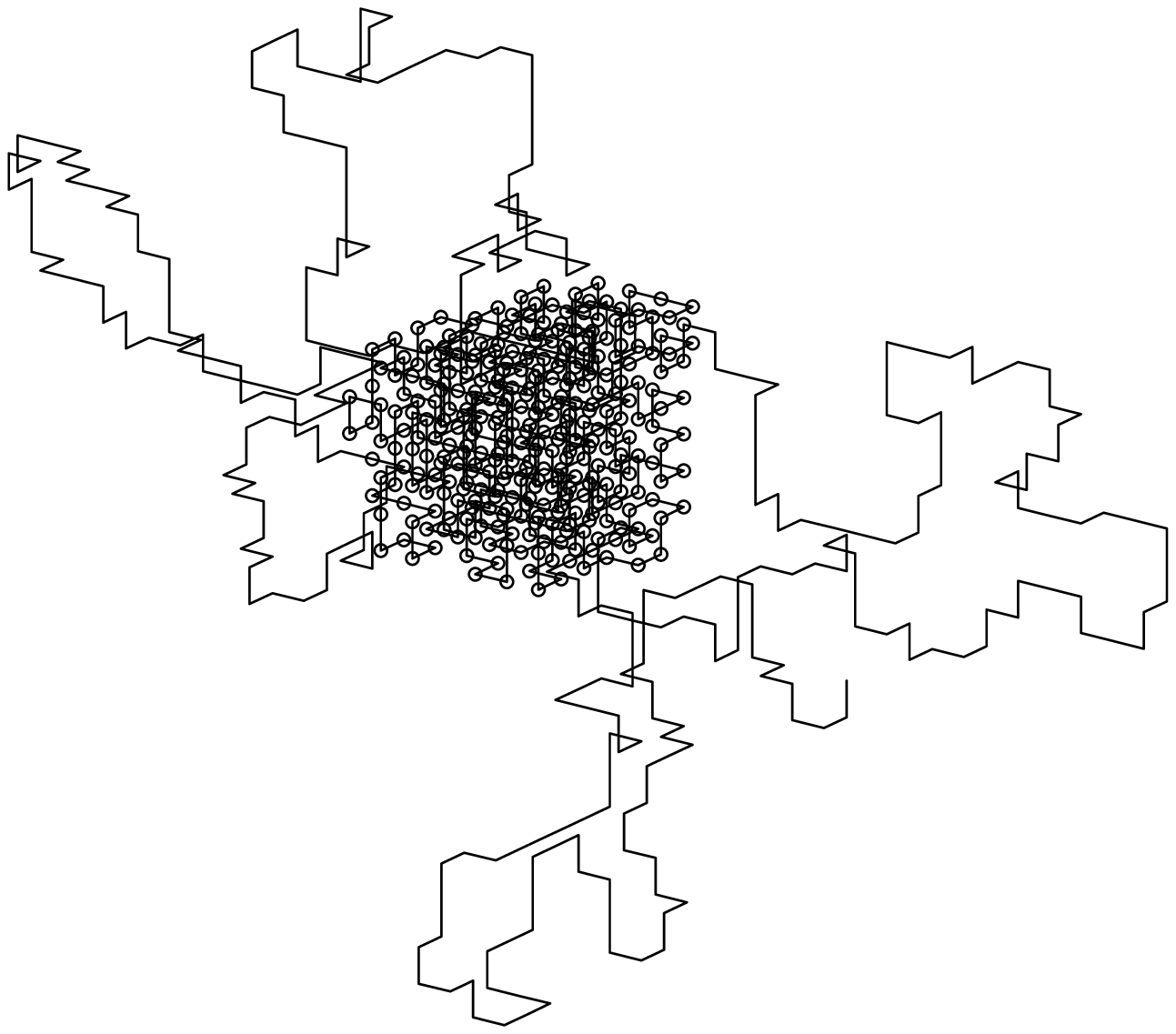}{0.8}
\vskip -70mm
\label{figure3}
\end{figure}

\begin{figure}
\vskip -5mm
\inseps{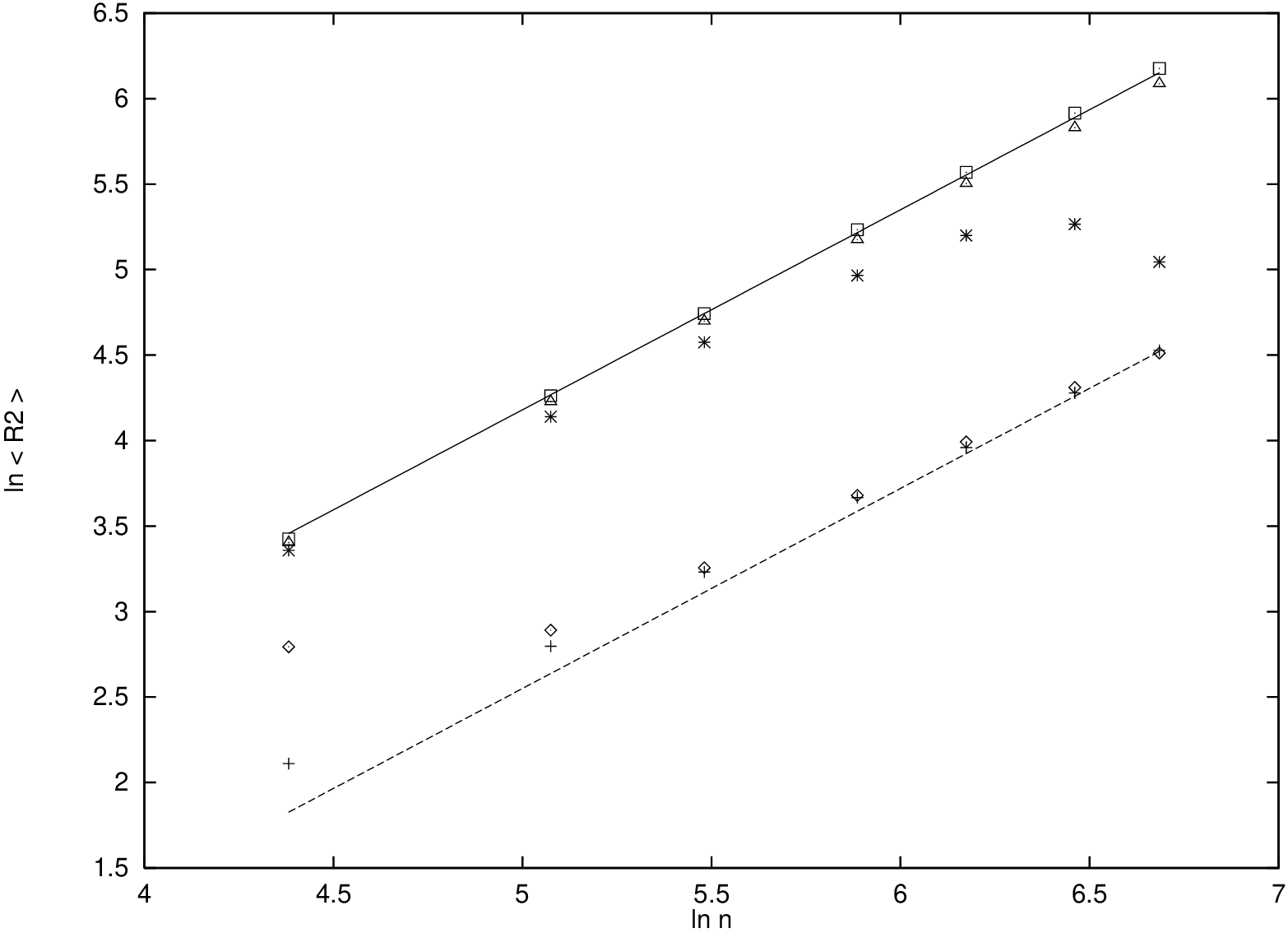}{0.7}
\vskip -10mm
\label{figure}
\end{figure}

\begin{figure}
\vskip -5mm
\inseps{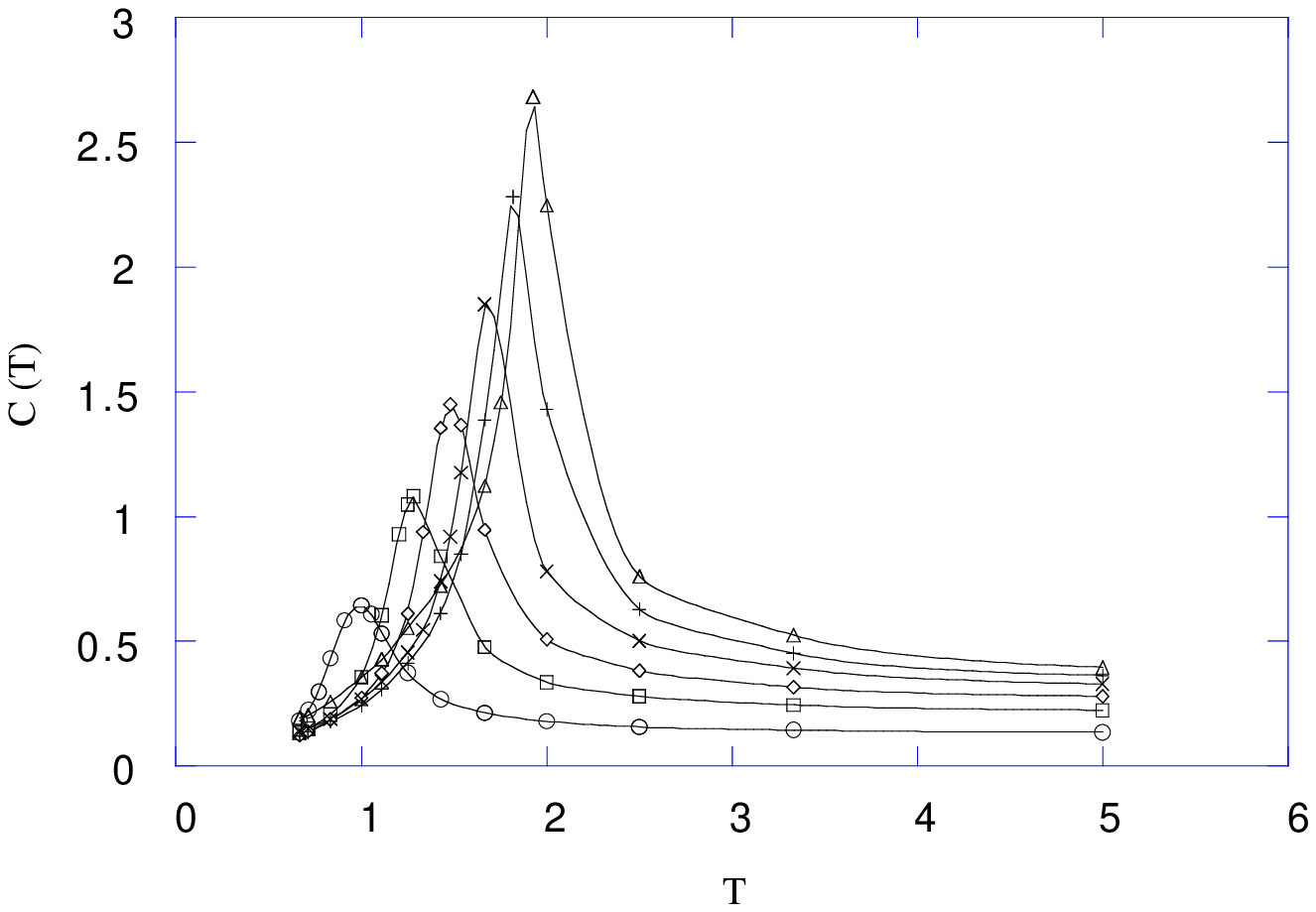}{1.2}
\vskip -10mm
\label{figure5}
\end{figure}

\begin{figure}
\vskip -5mm
\inseps{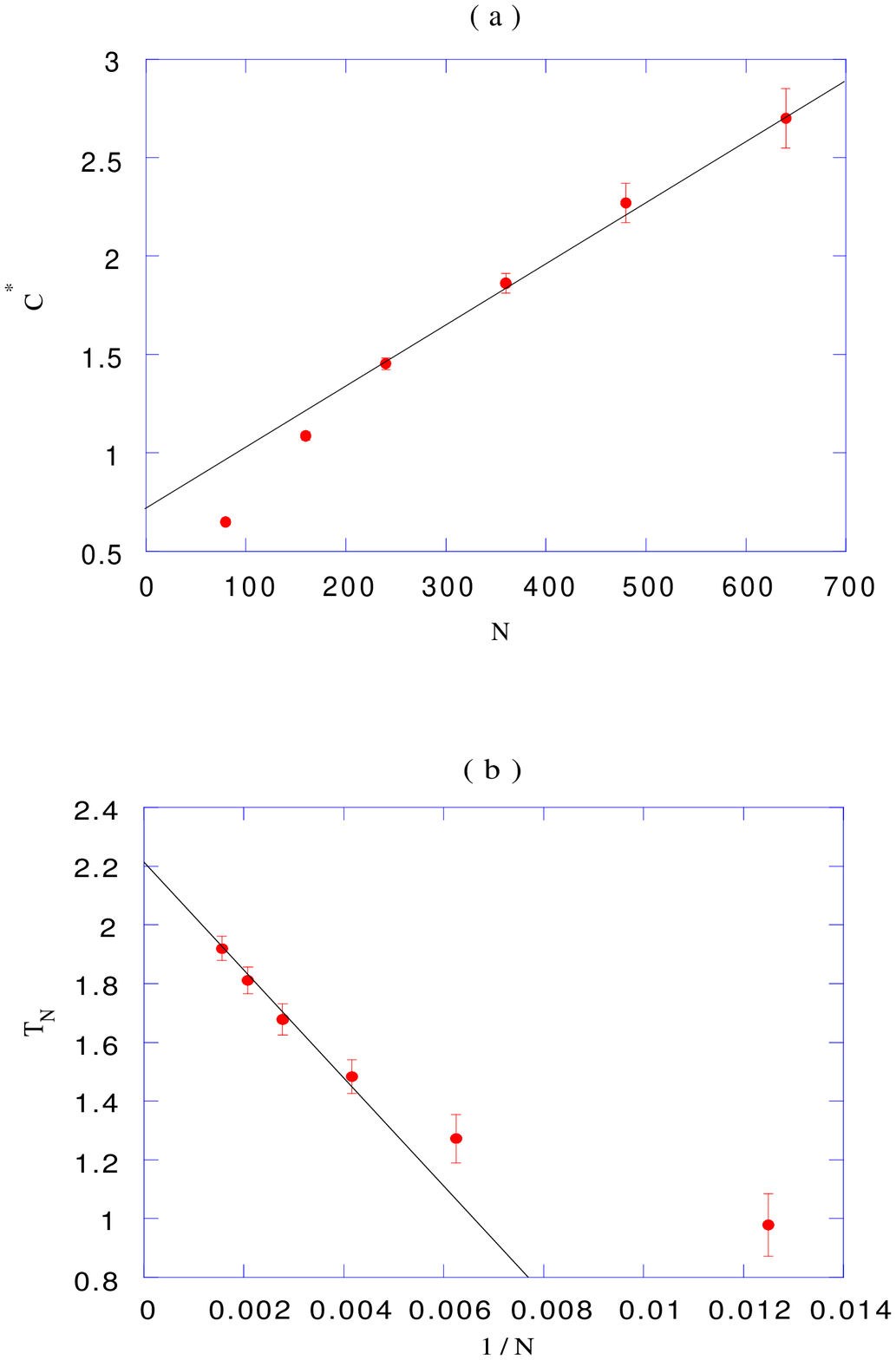}{0.75}
\vskip -10mm
\label{figure6}
\end{figure}


\begin{figure}
\phantom{.}
\vskip -10mm 
\inseps{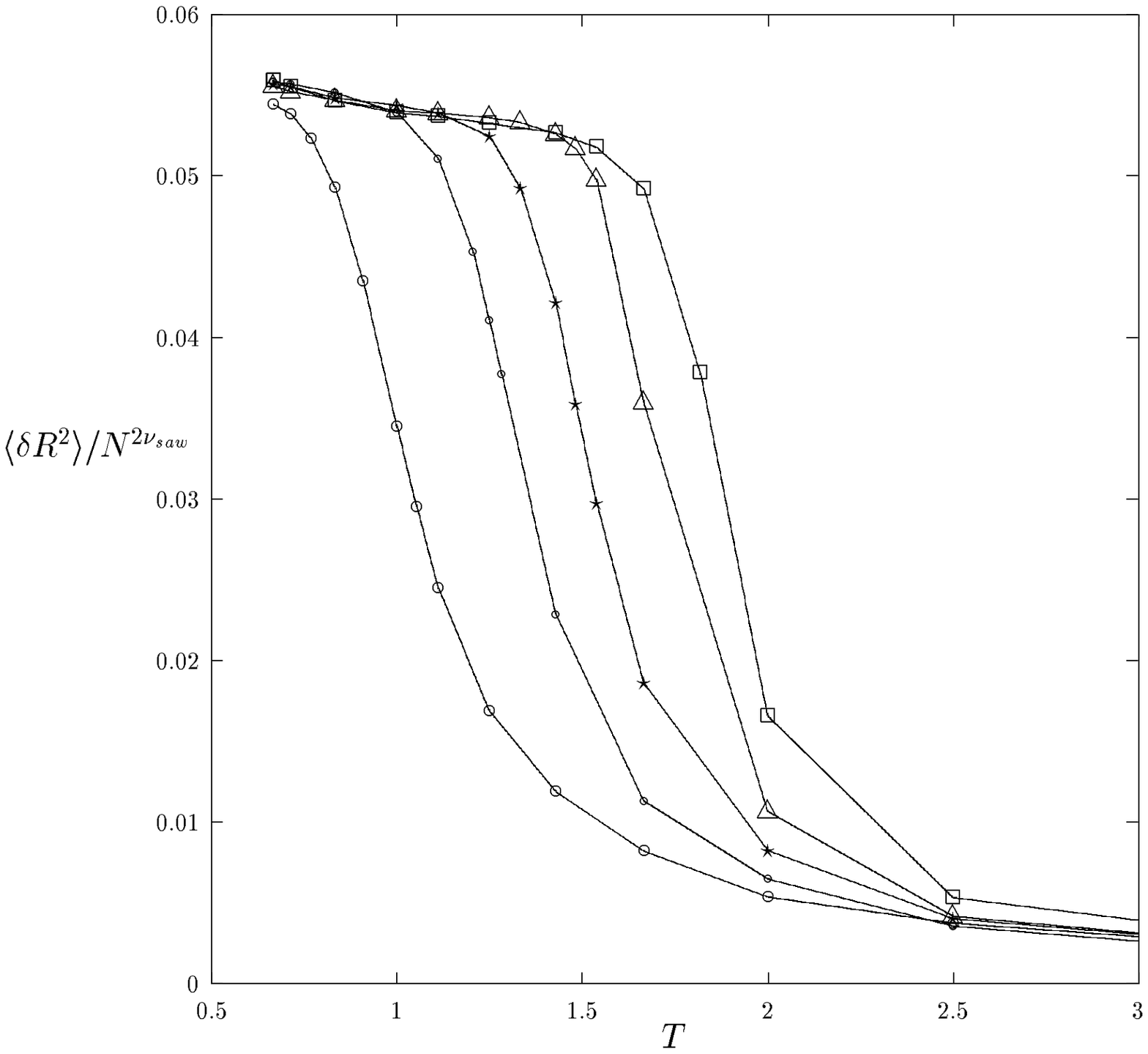}{0.9}
\vskip -50mm
\label{figure7}
\end{figure}

\begin{figure}
\vskip -5mm
\inseps{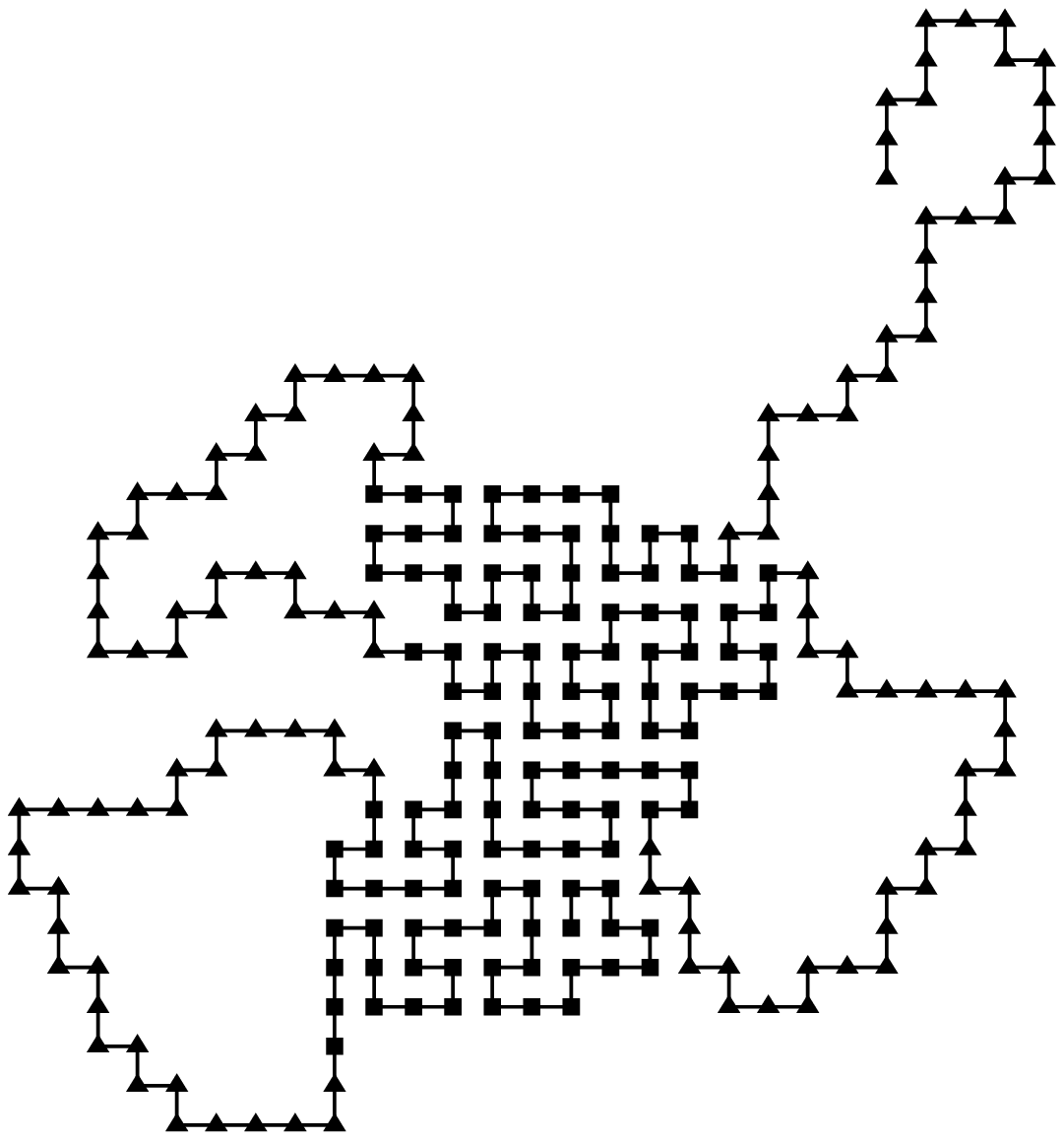}{0.9}
\vskip -10mm
\label{figure8}
\end{figure}

\begin{figure}
\vskip -5mm
\inseps{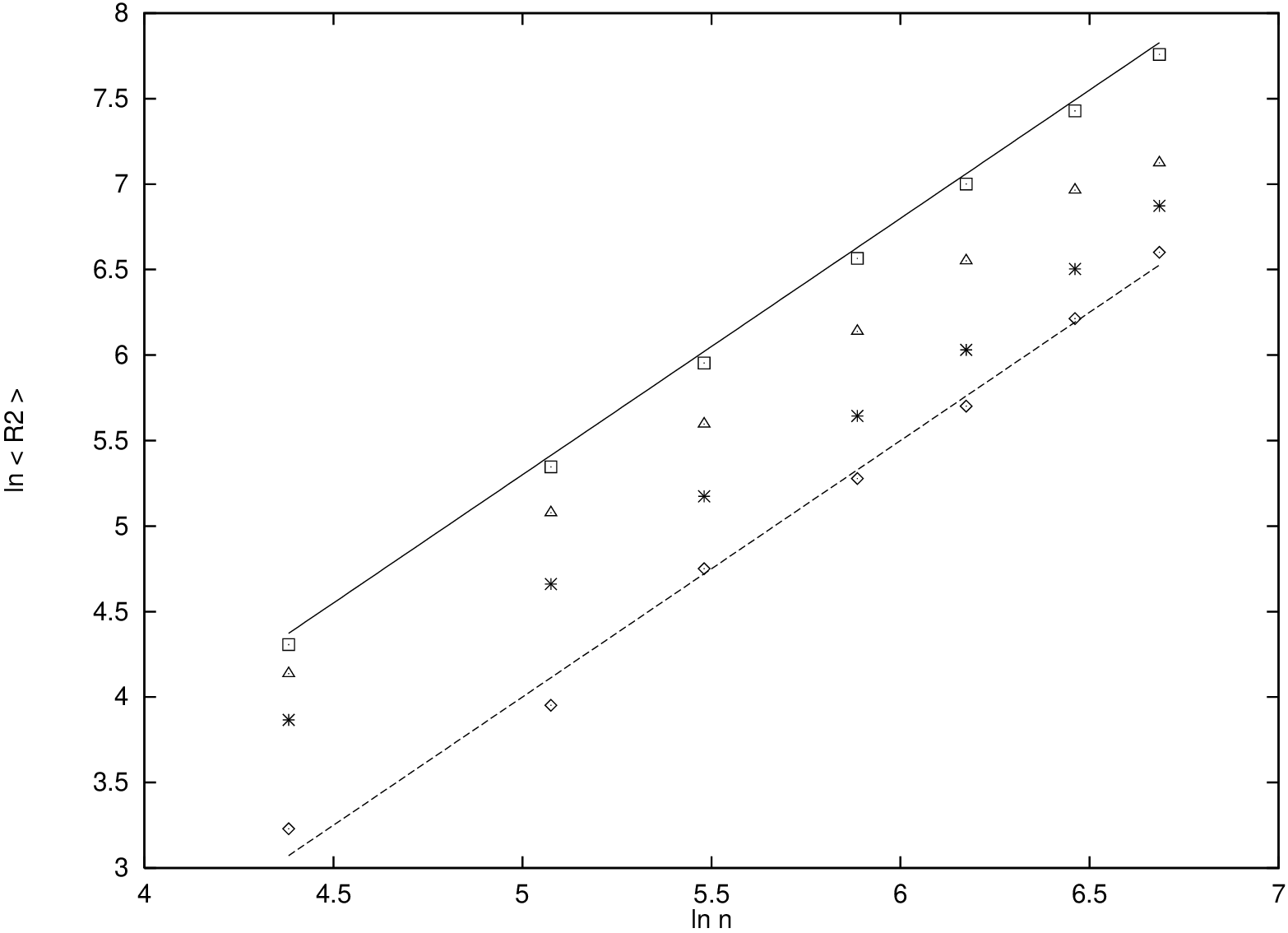}{0.7}
\vskip -10mm
\label{figure9}
\end{figure}

\begin{figure}
\vskip -5mm
\inseps{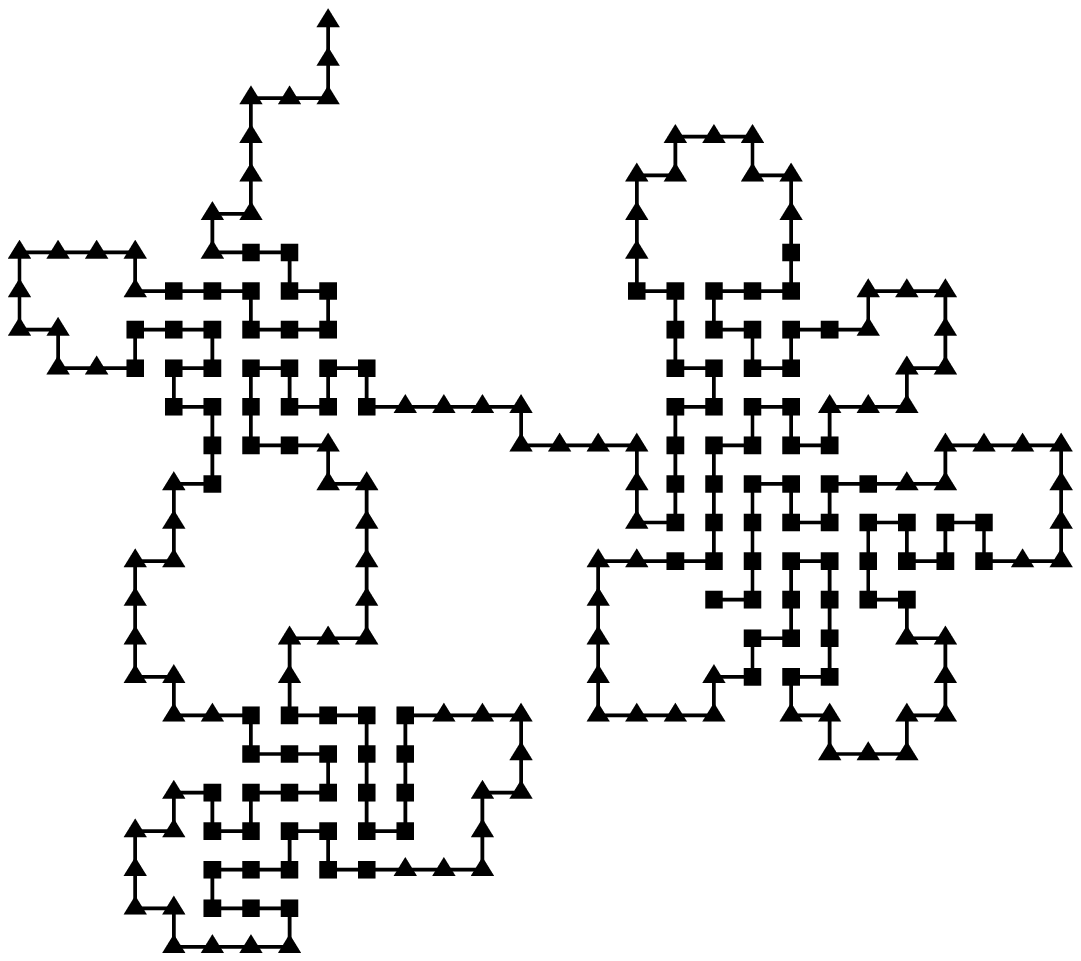}{0.9}
\vskip -10mm
\label{figure10}
\end{figure}

\begin{figure}
\vskip -5mm
\inseps{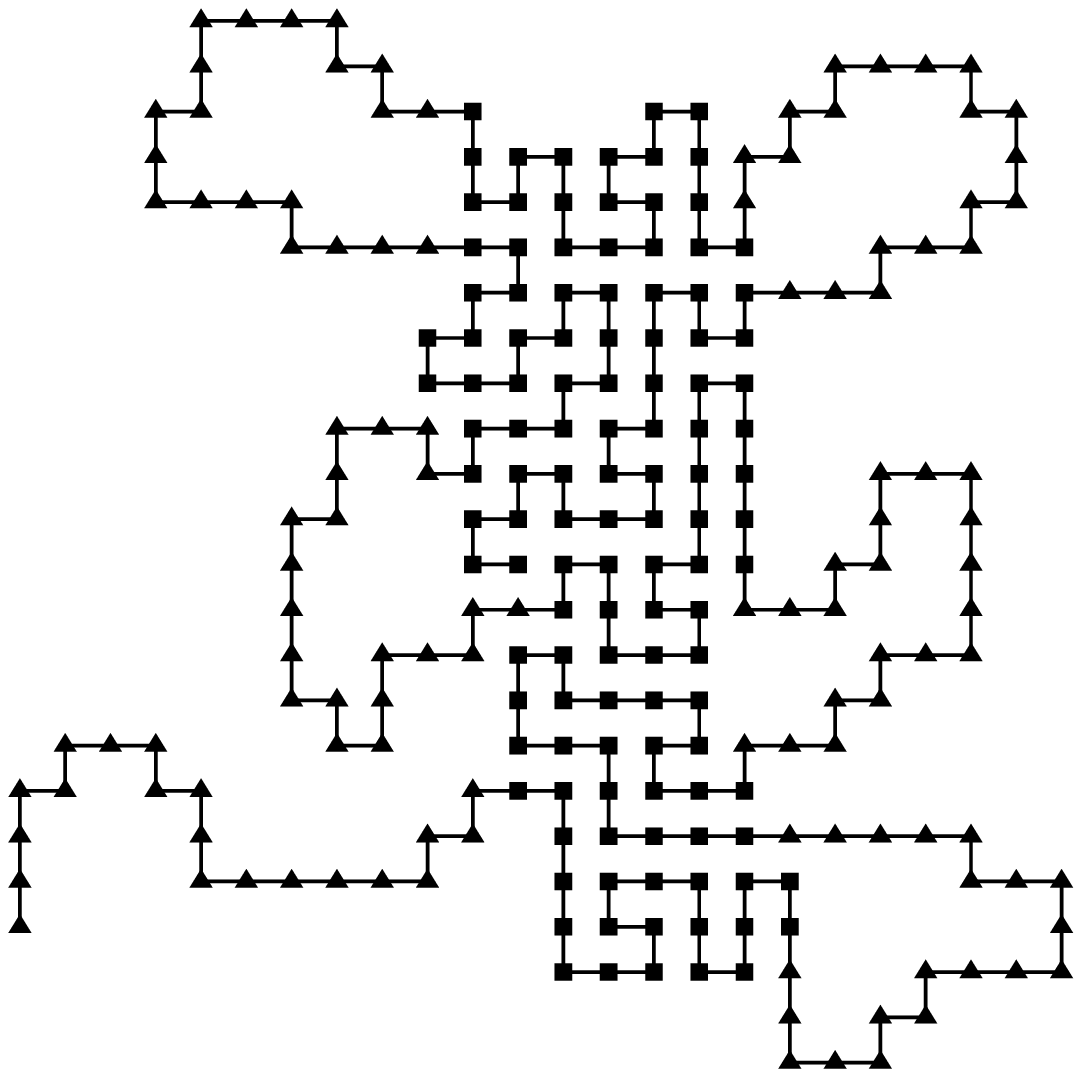}{0.9}
\vskip -10mm
\label{figure11}
\end{figure}

\begin{figure}
\vskip -5mm
\inseps{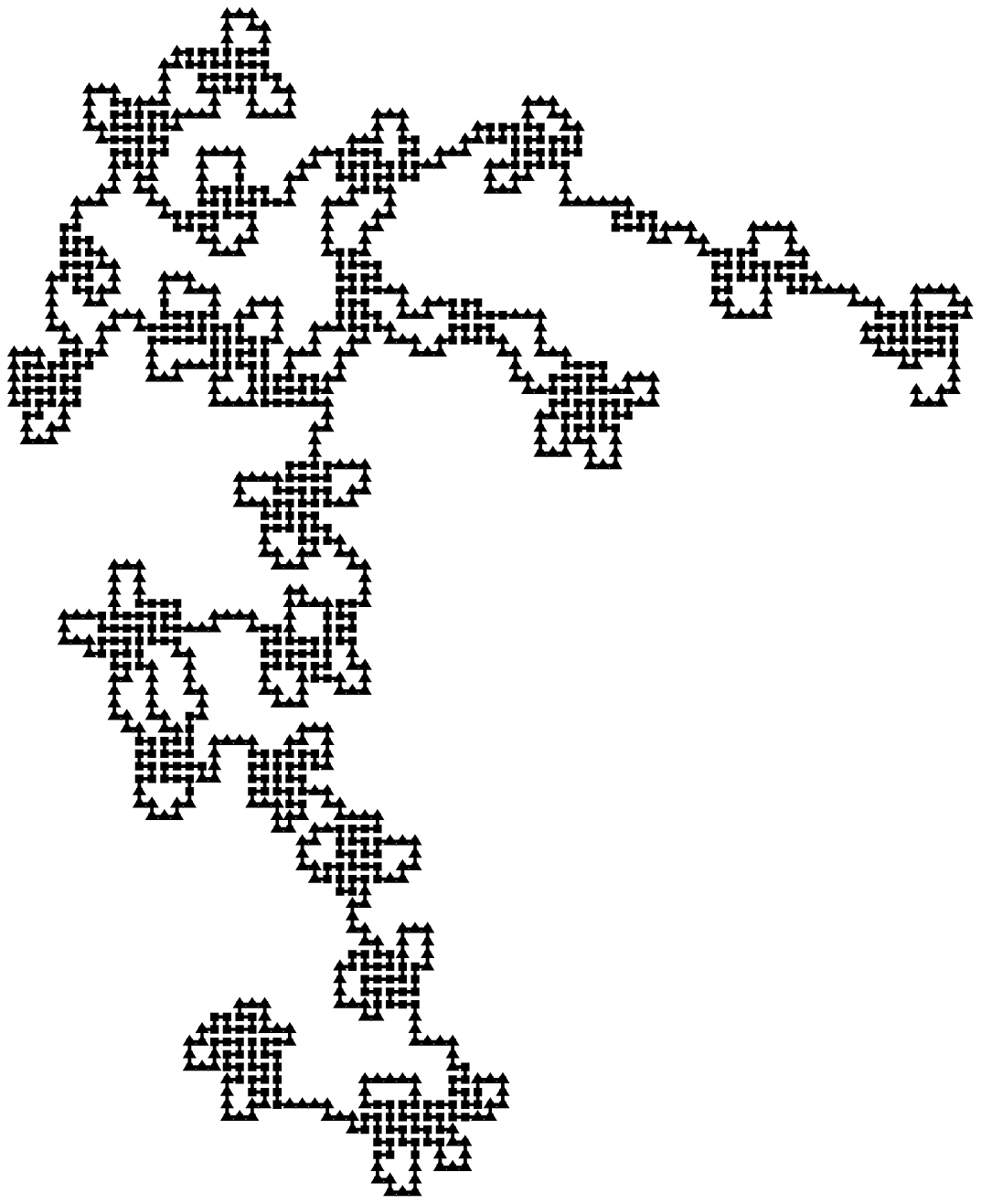}{0.9}
\vskip -10mm
\label{figure12}
\end{figure}

\begin{figure}
\vskip -5mm
\inseps{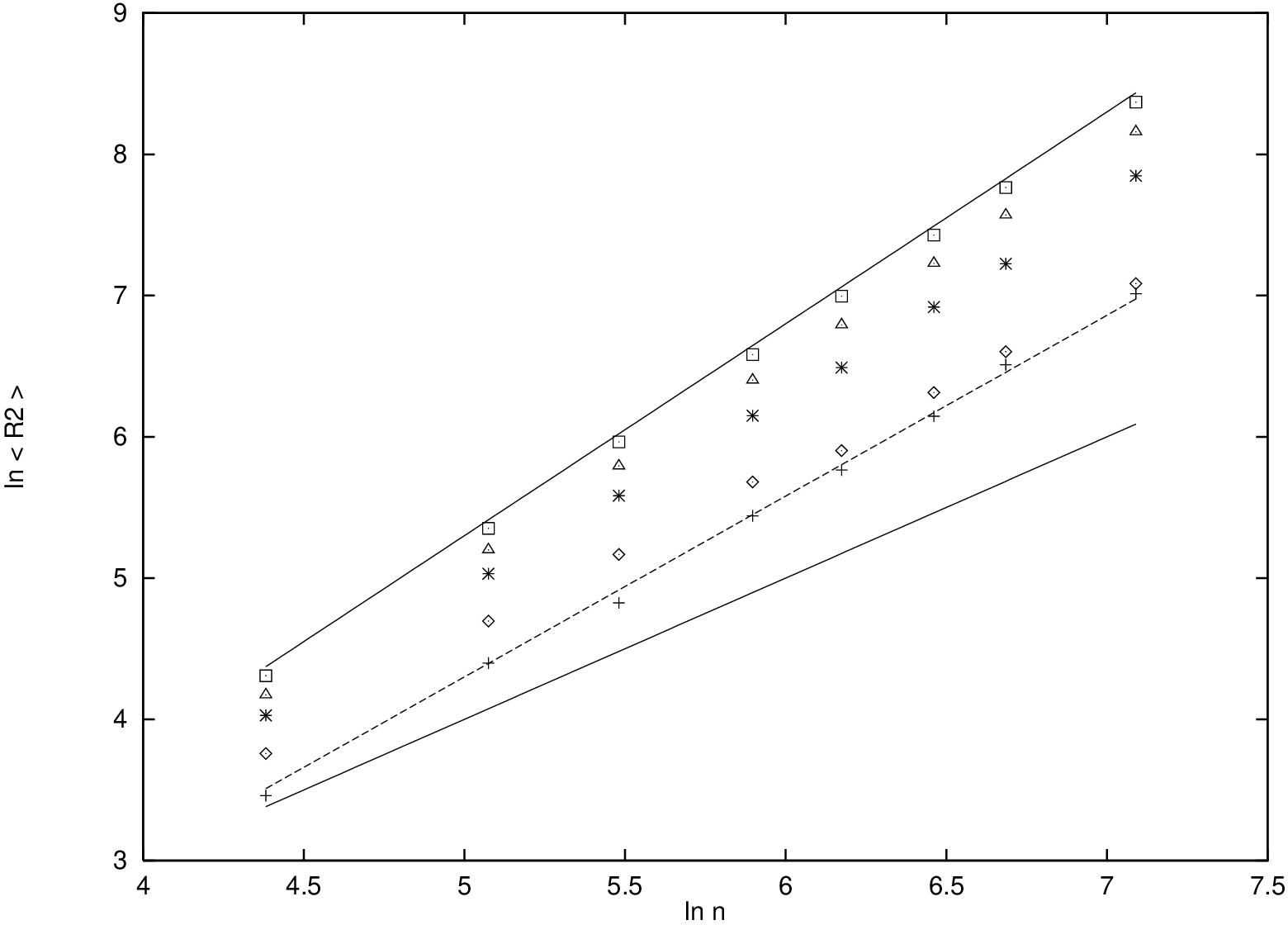}{0.7}
\vskip -10mm
\label{figure13}
\end{figure}

\begin{figure}
\vskip -1.truecm
\inseps{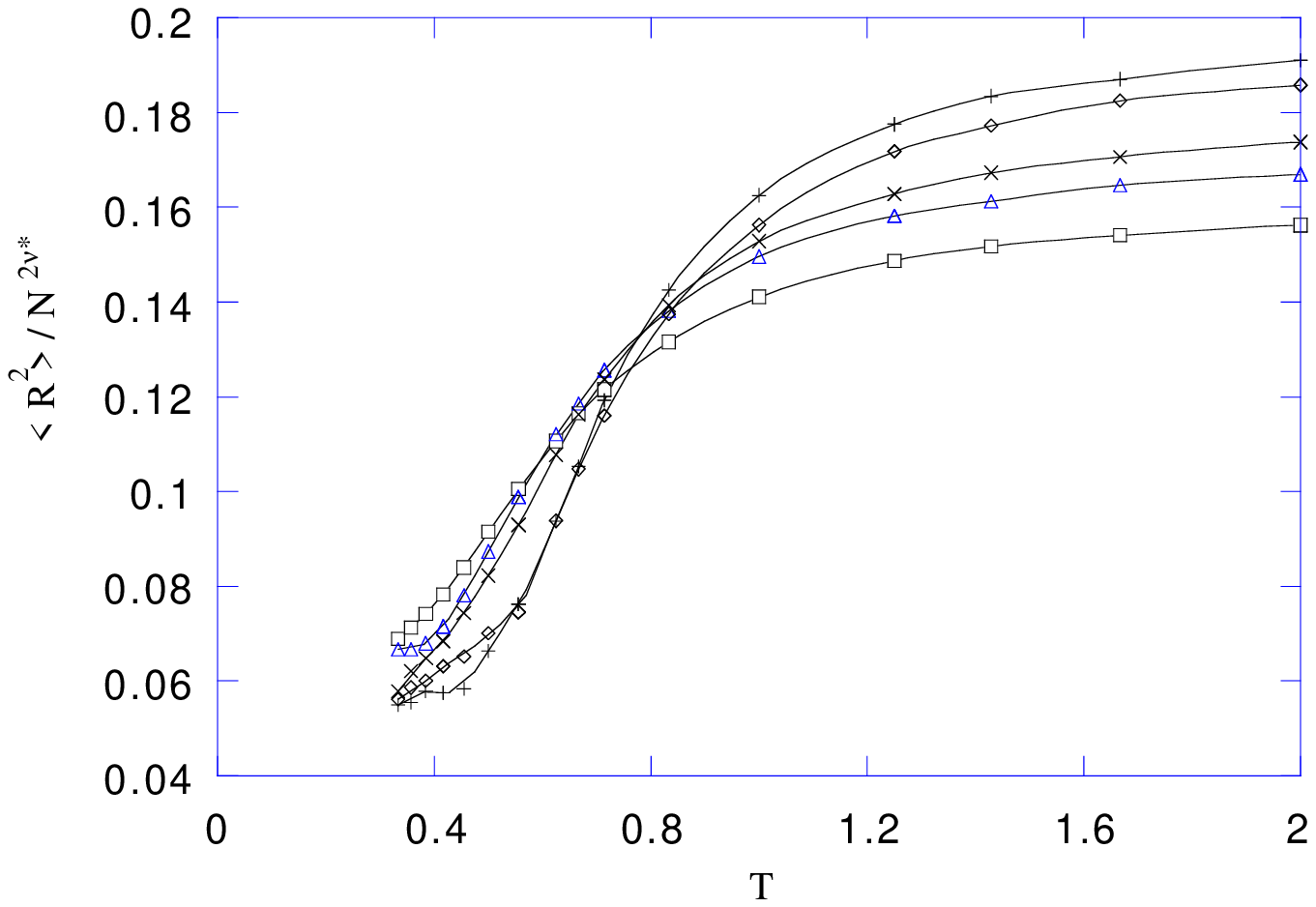}{1.2}
\label{figure14}
\end{figure}

\begin{figure}
\vskip -1.truecm
\inseps{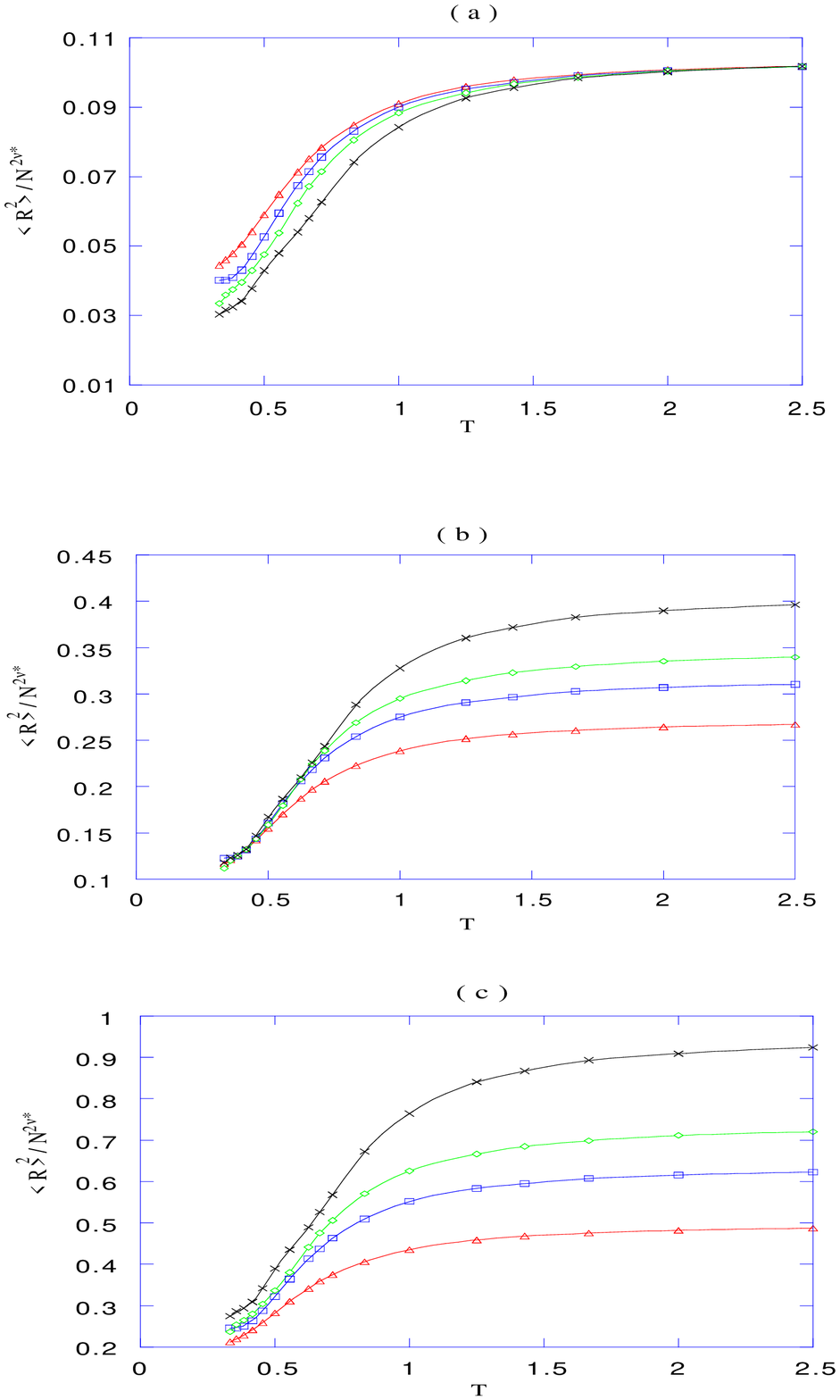}{0.75}
\label{figure15}
\end{figure}

\begin{figure}
\vskip -5mm
\inseps{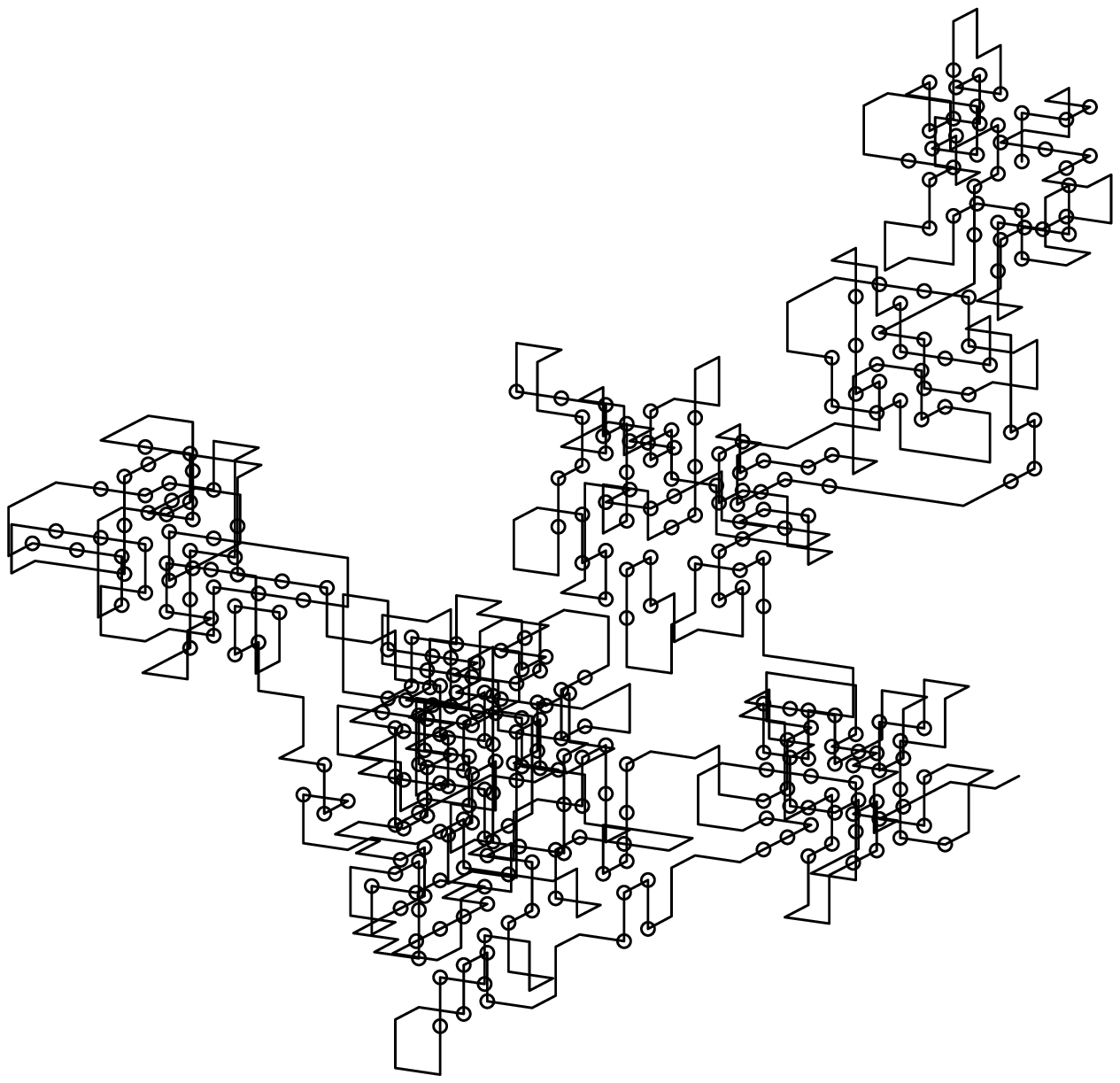}{0.8}
\vskip -50mm
\label{figure16}
\end{figure}

\begin{figure}
\vskip -50mm
\inseps{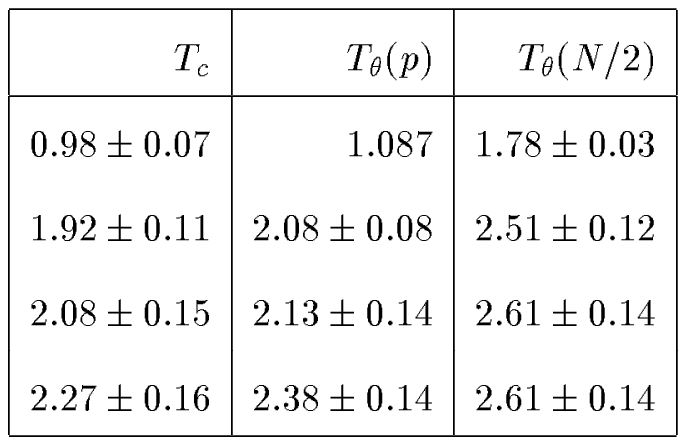}{1.0}
\end{figure}

\end{document}